\documentclass[10pt,conference]{IEEEtran}
\usepackage{cite}
\usepackage{amsmath,amssymb,amsfonts}
\usepackage{graphicx}
\usepackage{textcomp}
\usepackage{xcolor}
\usepackage[hyphens]{url}
\usepackage{booktabs}
\usepackage{multirow}
\usepackage{tikz}
\usetikzlibrary{shapes.geometric,calc}
\usepackage{pgfplots}
\pgfplotsset{compat=1.16}
\usepgfplotslibrary{groupplots,statistics}

\definecolor{figblue}{HTML}{3A6FB0}
\definecolor{figred}{HTML}{D9564E}
\definecolor{figgray}{HTML}{666666}
\definecolor{sev1}{HTML}{F5F1EA}
\definecolor{sev2}{HTML}{F0DCC8}
\definecolor{sev3}{HTML}{EAC3A3}
\definecolor{sev4}{HTML}{E2A57E}
\definecolor{sev5}{HTML}{D5825C}
\definecolor{sev6}{HTML}{C25E41}
\definecolor{sev7}{HTML}{A03B2B}
\definecolor{sev8}{HTML}{6E211A}
\pgfplotsset{hpcafig/.style={
  tick label style={font=\scriptsize},
  label style={font=\scriptsize},
  title style={font=\scriptsize},
  legend style={font=\scriptsize, draw=none, fill=none},
  legend cell align=left,
  ymajorgrids, grid style={gray!22, very thin},
  axis line style={gray!70, semithick},
  tick style={gray!70, semithick},
  scaled ticks=false,
  every axis plot/.append style={line width=0.9pt},
}}
\usepackage[hidelinks]{hyperref}

\title{Beyond Static Policies: Dynamic Selection Among Modern Microarchitectural Policies}

\author{
  \IEEEauthorblockN{
    Yanxin Zhang\IEEEauthorrefmark{1}\IEEEauthorrefmark{2},
    Ian McDougall\IEEEauthorrefmark{2},
    Junnan Li\IEEEauthorrefmark{2},\\
    Shayne Wadle\IEEEauthorrefmark{2},
    Vikas Singh\IEEEauthorrefmark{2},
    Karthikeyan Sankaralingam\IEEEauthorrefmark{1}\IEEEauthorrefmark{2}
  }
  \IEEEauthorblockA{
    \IEEEauthorrefmark{1}NVIDIA \qquad
    \IEEEauthorrefmark{2}UW--Madison\\
    \texttt{yanxinz@nvidia.com, imcdougall@wisc.edu, jli2786@wisc.edu}\\
    \texttt{swadle@cs.wisc.edu, vsingh@biostat.wisc.edu, karu@cs.wisc.edu}
  }
}

\begin{document}
\maketitle

\begin{abstract}
Modern processors gain performance from interacting policies: prefetchers, predictors, replacement rules, and schedulers. These policies are often evaluated one at a time, yet a policy that wins in one stack may lose in another. To study these effects, we present the first systematic composition study of two L1D prefetchers, two L1I prefetchers, and two L2 replacement policies across 490 phases from 49 SPEC CPU2006 and SPEC CPU 2017 traces. We define the best global static policy (BGSP) by phase-level oracle-win frequency. Gaze/Entangling/Mockingjay is the BGSP, winning 33.47\% of phases, yet it remains 1.33\% below the phase oracle on average, with 52 phases across eight benchmarks losing more than 2.5\%. The opportunity is highly compressible: a Berti/Gaze pair that changes only the L1D prefetcher comes within 0.039\% aggregate IPC of the eight-configuration oracle, reducing runtime control to one bit per 200K-instruction window. Given that one-bit interface, we frame selector design as an information problem: what can hardware know before choosing? We evaluate selectors that use only chosen-policy IPC, selectors that passively monitor the demand stream before either prefetcher changes cache state, and an ideal counterfactual observer that exposes the inactive-policy winner signal. The main practical result is that both executed-performance feedback and passive demand monitoring techniques capture much of the two-policy opportunity, recovering 62.4\% to 73.4\% of the pairwise oracle gap without executing or emulating the inactive prefetcher. The counterfactual study shows that inactive-policy observation must be nearly exact and available within one window to improve on executed-performance or passive demand monitoring. These results suggest a general method for adapting among microarchitectural policies as an additional pathway for processor improvement, distinct from structural resizing.
\end{abstract}

\section{Introduction}
\label{sec:introduction}

Modern processors rely on several interacting policies within the
memory hierarchy, including data and instruction prefetchers and cache
replacement. New policies are typically evaluated one component at a time: a new
data prefetcher is compared against alternatives while the surrounding
memory-system policies stay fixed. This isolates the benefit of a design choice,
but it hides the fact that these policies affect one another when used together.
A data prefetcher changes the requests seen by lower cache levels, instruction
prefetching changes the timing and overlap of memory accesses, and replacement
determines which prefetched lines remain available when they are needed. The
policy that performs best for one component can therefore depend on the others it
is paired with. This raises a question: when multiple strong modern policies are
available, is there one combination that works well throughout execution, or does
the best combination vary enough across workloads and over time to make runtime
adaptation worthwhile?

We study these questions by jointly evaluating two L1D prefetchers, two L1I
prefetchers, and two L2 replacement policies, yielding eight complete
memory-system combinations. We evaluate them across 490 execution phases from 49
SPEC CPU2006 and SPEC CPU2017 traces, comparing each fixed combination against an
oracle that picks the best-performing combination per phase. To our knowledge,
this is the first systematic study of this kind. At first glance, runtime
adaptation looks unnecessary: the best fixed choice, which we call the
\textit{best global static policy} (BGSP), Gaze/Entangling/Mockingjay, is only
1.33\% behind the oracle on average. But collapsing the measurements to an
average discards meaningful distributional information. Across the 490 phases, 52
lose more than 2.5\%, 39 more than 5\%, and 18 more than 10\% relative to the
oracle, and these losses are spread across the benchmark suite. A single fixed
choice thus looks strong in aggregate while leaving many per-application and
per-phase opportunities unexplored.

Exploiting these opportunities need not require choosing among all eight
combinations at runtime. An eight-way selector is costly: the processor must
support every candidate policy, manage its state, and can observe only the
combination that actually runs, never the seven alternatives. We therefore ask
whether most of the benefit survives with a much smaller set of choices, and it
does. On the 4,900 200K-instruction windows used for our runtime study, a pair
that fixes the L1I prefetcher and L2 replacement policy and varies only the L1D
prefetcher between Berti and Gaze comes within 0.039\% aggregate IPC of the full
eight-combination oracle. The design space thus collapses from eight combinations
to a single binary decision: whether to use Berti or Gaze.

Two choices still leave the question of how to decide between them, since the
processor observes the performance of the prefetcher currently running but not
what the other would have done. One option is to learn from each choice's
performance as it is tried. Using UCB, we improve mean loss to 0.407\% and
accuracy to 68.41\%, recovering 73.4\% of the maximum available runtime
gains---but only by occasionally trying a choice that may be worse. We therefore
ask whether the decision can instead be made from program behavior that is
visible regardless of which prefetcher is active. Using six demand-side features
and a depth-3 decision tree, mean loss is slightly higher than UCB at 0.465\% on
held-out traces, but accuracy improves to 72.6\%, recovering 62.4\% of the
maximum available runtime gains.

Finally, we ask whether the selector could do better if it knew which prefetcher
performed better in the recent past. Augmenting it with the identity of the
better prefetcher from the preceding window, perfect and immediately available
information drops mean loss from 0.465\% to 0.320\%. But this advantage mostly
vanishes once the information is delayed by one additional window or is wrong just
10\% of the time. Unobserved alternative-policy performance thus offers limited
value unless it can be obtained accurately and promptly, so we adopt the passive
demand-based selector, which relies only on information available regardless of
the active prefetcher and requires no explicit exploration.

This paper makes three contributions:
\begin{itemize}
\item We present, to our knowledge, the first systematic composition study of eight complete L1D$\times$L1I$\times$L2 configurations. The BGSP is close to the oracle on average, yet leaves a costly phase-level tail spread across eight of 49 benchmarks.
\item We show that oracle opportunity is highly compressible. A Berti/Gaze pair with fixed L1I prefetching and L2 replacement comes within 0.039\% aggregate IPC of the full oracle, reducing cross-stack adaptation to a one-bit L1D-prefetcher decision.
\item We evaluate runtime selectors by the information available to the PSB. Executed-performance feedback and passive demand monitoring capture much of the two-policy opportunity, recovering 62.4\% to 73.4\% of the pairwise oracle gap; idealized counterfactual information provides only a small, delay-sensitive upper-bound improvement.
\end{itemize}

Sections~\ref{sec:oracle_methodology} and~\ref{sec:oracle_results} describe the composition study and phase-oracle results. Section~\ref{sec:compression} compresses the action space to a Berti/Gaze pair. Sections~\ref{sec:selector_design} and~\ref{sec:evaluation} define the PSB interface, evaluate selector information sources, and end with the counterfactual upper bound in Section~\ref{sec:ipu}.

\section{Experimental Setup}
\label{sec:oracle_methodology}

\subsection{The Composition Design Space}
A modern out-of-order core overlaps instruction fetch, data access, and lower-level cache service, so policies which operate at different levels of the memory system do not act independently. We therefore study three policy choices jointly: L1I prefetching, L1D prefetching, and L2 replacement. Each memory-system combination contains one policy for each of these three axes. Since these policies are typically evaluated one component at a time, the policy that performs best for one component need not remain best when the surrounding policy choices change. \emph{Our first question is simple: when strong policies for these three components are evaluated together, does one combination remain best throughout execution? Or do different parts of execution prefer different combinations?}

\subsection{Candidate Policies}

For each of the three components, we select two recent, competitive policies with different underlying mechanisms. This ensures that the study compares meaningful/strong alternatives. For L1D prefetching, Berti~\cite{navarro2022berti} tracks address deltas associated with individual instructions and selects deltas that can arrive in time. And  Gaze~\cite{chen2025gaze} identifies recurring spatial-region footprints through temporal correlation.

For L1I prefetching, Entangling~\cite{ros2021cost} links instruction-cache lines that tend to be requested together, while BARCA~\cite{gratz2020barca} uses branch-agnostic region search to improve instruction delivery. For L2 replacement, Mockingjay~\cite{shah2022effective} and PACIPV~\cite{mostofi2025light} both predict future reuse, but use different state and prediction mechanisms. These six policies together give us the $2\times 2 \times 2$ design space of eight memory-system combinations, with one L1D prefetcher, one L1I prefetcher, and one L2 replacement policy in each.

\subsection{System Configuration and Simulator}

Every configuration is measured on the same canonical 6-wide out-of-order processor shown in Table~\ref{tab:config}. All eight configurations are implemented within a common ChampSim codebase~\cite{gober2022championship}, sharing the core, hierarchy, memory model, trace reader, warmup, and measurement settings. 
Therefore, across the eight configurations, the only intended differences are the L1I prefetcher, L1D prefetcher, and L2 replacement policy being evaluated, 
which avoids simulator or configuration drift from being misattributed to other factors.

\begin{table}[tb]
\centering
\caption{Simulated system configuration.}
\label{tab:config}
\begin{tabular}{@{}p{0.35\columnwidth}p{0.57\columnwidth}@{}}
\toprule
Parameter & Value \\
\midrule
Core & 6-wide OOO \\
Pipeline & 6-wide fetch/decode/dispatch; 4-wide execute; 5-wide retire \\
ROB & 352 entries \\
LSQ & 128-entry LQ, 72-entry SQ \\
Branch predictor & TAGE-SC-L \\
\midrule
L1I & 32\,KB, 8-way, 4-cycle \\
L1I prefetcher & \emph{(varied)} Entangling or BARCA \\
L1D & 48\,KB, 12-way, 5-cycle \\
L1D prefetcher & \emph{(varied)} Berti or Gaze \\
\midrule
L2 & 512\,KB, 8-way, 10-cycle \\
L2 replacement & \emph{(varied)} Mockingjay or PACIPV \\
LLC & 2\,MB, 16-way, 20-cycle \\
Memory & Single-channel 3200\,MT/s DRAM \\
\bottomrule
\end{tabular}
\end{table}


\subsection{Workloads}

 We use 49 public ChampSim traces drawn from the SPEC CPU2006 and SPEC CPU 2017 benchmark suites and released with the DPC-3 trace set~\cite{gober2022championship}.
 So, the workload set is standard, public, and reproducible.
 Every configuration uses identical instruction counts, phase boundaries, and window boundaries (defined shortly). An IP-stride/LRU system provides a reference point for calibrating how much of the available improvement modern policies already capture, but it is not a selector candidate; all composition and selector comparisons among candidates use the modern configurations.

\subsection{Phases, Windows, and Metrics}

A single number per application does not show whether different program regions prefer different policy combinations, or how much performance is available when they do. The study therefore evaluates complete configurations per phase. A \emph{phase} is a 20M-instruction interval used to measure how much headroom exists: long intervals amortize metadata warmup and suppress short-interval IPC noise that could otherwise swamp small differences among configurations. A \emph{window} is a 200K-instruction interval used later to evaluate whether a controller can capture that headroom: short intervals provide enough chronological decisions for replay and expose the noise and responsiveness constraints a runtime selector must handle.

For each phase $\phi$, all eight configurations execute the same instruction interval under identical simulation settings. The phase-optimal oracle is the maximum IPC over the eight-configuration set $C$:
\begin{equation}
IPC_{\mathit{oracle}}(\phi)=\max_{c\in C} IPC(c,\phi).
\end{equation}
The oracle is computed after all eight runs complete, ignores switching cost, and performs no prediction, so it is an opportunity bound, not a hardware proposal. Its purpose is to measure the performance left by any fixed configuration choice, not to define an implementable controller.

A real machine runs one configuration during a phase, so IPC loss is the shortfall of that configuration against the phase oracle:
\begin{equation}
Loss(c,\phi)=
\frac{IPC_{\mathit{oracle}}(\phi)-IPC(c,\phi)}
     {IPC_{\mathit{oracle}}(\phi)}\times100\%.
\end{equation}
Mean loss averages this value over phases and is the primary metric because every phase contributes equally to the question of behavioral diversity. Aggregate loss sums IPC before normalizing,
\begin{equation}
Loss_{\mathit{agg}}(c)=
\frac{\sum_{\phi} IPC_{\mathit{oracle}}(\phi)-\sum_{\phi} IPC(c,\phi)}
     {\sum_{\phi} IPC_{\mathit{oracle}}(\phi)}\times100\%,
\end{equation}
and measures total-throughput impact when higher-IPC phases carry more weight.

The distribution of losses is as important as the mean. We report the fraction of phases above 0.5\%, 1\%, 2.5\%, 5\%, and 10\% loss because the tail is the effect a mean cannot express. Section~\ref{sec:oracle_results} uses these metrics to show where static composition leaves headroom; later sections keep the same phase/window naming discipline while narrowing the action space and evaluating runtime selection.

\section{Composition Results: Headroom and Candidate Reduction}
\label{sec:oracle_results}

This section starts from the eight complete configurations defined in Section~\ref{sec:oracle_methodology} and asks how much static selection leaves behind. We first examine the best global static policy in aggregate, then separate the same data by application to see whether the losses are concentrated or spread across the suite. We then compare all eight policies against the phase oracle and use the oracle choices to identify the smaller candidate set that later runtime selectors must choose between.

\subsection{Single-Policy Analysis: Per-Application Variability}

We define the \emph{best global static policy} (BGSP) as the configuration with the highest phase-level oracle-win frequency. Gaze/Entangling/Mockingjay satisfies this criterion, winning 33.47\% of the 490 phases. We use this configuration as the static reference throughout the coarse-phase analysis. The goal is to determine whether different applications, and different phases within an application, prefer alternatives strongly enough to justify building more than one configuration and choosing among them at runtime.

\begin{figure}[tb]
\centering
\begin{tikzpicture}
\begin{axis}[
  hpcafig,
  width=0.92\columnwidth, height=4.4cm,
  xmin=0, xmax=39,
  xtick={0,10,20,30},
  xlabel={Percent of phases (\%)},
  xlabel style={yshift=2pt},
  xmajorgrids, ymajorgrids=false,
  grid style={gray!22, very thin},
  symbolic y coords={{$>$10\%},{5--10\%},{2--5\%},{1--2\%},{0.5--1\%},{0.1--0.5\%},{0--0.1\%},{0\% (opt.)}},
  ytick={{$>$10\%},{5--10\%},{2--5\%},{1--2\%},{0.5--1\%},{0.1--0.5\%},{0--0.1\%},{0\% (opt.)}},
  ylabel={IPC-loss range},
  enlarge y limits=0.09,
  nodes near coords={\pgfmathprintnumber[fixed,precision=1,zerofill]{\pgfplotspointmeta}},
  every node near coord/.append style={font=\scriptsize, anchor=west, xshift=1pt},
  clip=false,
]
\addplot[xbar, bar shift=0pt, bar width=8pt, point meta=rawx, fill=sev1, draw=gray!80, line width=0.35pt] coordinates {(33.5,{0\% (opt.)})};
\addplot[xbar, bar shift=0pt, bar width=8pt, point meta=rawx, fill=sev2, draw=gray!80, line width=0.35pt] coordinates {(24.9,{0--0.1\%})};
\addplot[xbar, bar shift=0pt, bar width=8pt, point meta=rawx, fill=sev3, draw=gray!80, line width=0.35pt] coordinates {(16.9,{0.1--0.5\%})};
\addplot[xbar, bar shift=0pt, bar width=8pt, point meta=rawx, fill=sev4, draw=gray!80, line width=0.35pt] coordinates {(5.3,{0.5--1\%})};
\addplot[xbar, bar shift=0pt, bar width=8pt, point meta=rawx, fill=sev5, draw=gray!80, line width=0.35pt] coordinates {(7.8,{1--2\%})};
\addplot[xbar, bar shift=0pt, bar width=8pt, point meta=rawx, fill=sev6, draw=gray!80, line width=0.35pt] coordinates {(3.7,{2--5\%})};
\addplot[xbar, bar shift=0pt, bar width=8pt, point meta=rawx, fill=sev7, draw=gray!80, line width=0.35pt] coordinates {(4.3,{5--10\%})};
\addplot[xbar, bar shift=0pt, bar width=8pt, point meta=rawx, fill=sev8, draw=gray!80, line width=0.35pt] coordinates {(3.7,{$>$10\%})};
\end{axis}
\end{tikzpicture}
\caption{Global IPC-loss distribution of the BGSP, Gaze/Entangling/Mockingjay. Most phases are close to the oracle, but a nontrivial tail remains.}
\label{fig:static_loss_single}
\label{fig:static_loss_ranges}
\end{figure}
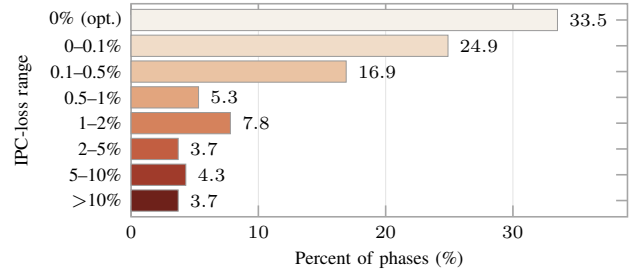

\emph{Aggregate behavior.} We first analyze how policy effectiveness varies across execution phases in bulk. Figure~\ref{fig:static_loss_single} shows the IPC-loss distribution of the BGSP across all 490 phases. The mean loss is 1.33\%, but that average hides the distributional information: IPC loss exceeds 1\% in 95 phases (19.4\%), 2.5\% in 52 (10.6\%), 5\% in 39 (8.0\%), and 10\% in 18 (3.7\%). Static selection therefore works well for many phases, yet it leaves a visible tail where another complete configuration would have delivered substantially higher IPC. A 5\% IPC gap is large enough to change design conclusions, so the upper tail is a meaningful signal.

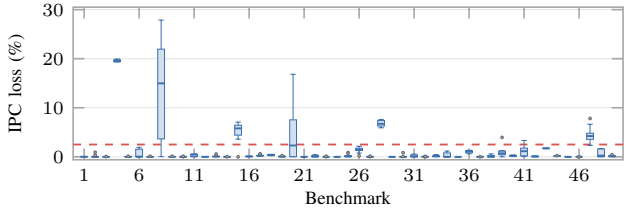
\begin{figure}[tb]
\centering
\begin{tikzpicture}
\begin{axis}[
  hpcafig,
  width=\columnwidth, height=3.6cm,
  ylabel={IPC loss (\%)},
  ymin=-0.6, ymax=30.5,
  ytick={0,10,20,30},
  xmin=0, xmax=50,
  xtick={1,6,11,16,21,26,31,36,41,46},
  xlabel={Benchmark},
  xlabel style={yshift=3pt},
  boxplot/draw direction=y,
  boxplot/box extend=0.62,
  every boxplot/.style={
    draw=figblue, fill=figblue!22, line width=0.35pt, solid,
    mark=*, mark size=0.5pt,
    mark options={fill=gray!60, draw=figgray, line width=0.2pt, solid},
    /pgfplots/boxplot/every median/.style={figblue, line width=0.7pt, solid},
  },
]
\addplot[figred, dashed, line width=0.7pt, forget plot] coordinates {(0,2.5) (50,2.5)};
\addplot+[boxplot prepared={draw position=1, lower whisker=0, lower quartile=0.0012, median=0.0156, upper quartile=0.0457, upper whisker=0.0629}] coordinates {};
\addplot+[boxplot prepared={draw position=2, lower whisker=0, lower quartile=0, median=0, upper quartile=0.0134, upper whisker=0.0179}] coordinates {(2,0.9278) (2,0.3516)};
\addplot+[boxplot prepared={draw position=3, lower whisker=0, lower quartile=0, median=0.0014, upper quartile=0.0078, upper whisker=0.012}] coordinates {(3,0.0612)};
\addplot+[draw=figred, fill=figred!28, boxplot prepared={draw position=4, lower whisker=19.3127, lower quartile=19.3997, median=19.5915, upper quartile=19.7669, upper whisker=19.9462}] coordinates {};
\addplot+[boxplot prepared={draw position=5, lower whisker=0, lower quartile=0.0002, median=0.003, upper quartile=0.0247, upper whisker=0.0518}] coordinates {(5,0.116)};
\addplot+[boxplot prepared={draw position=6, lower whisker=0, lower quartile=0, median=0.0307, upper quartile=1.5433, upper whisker=1.9346}] coordinates {};
\addplot+[boxplot prepared={draw position=7, lower whisker=0, lower quartile=0, median=0, upper quartile=0.0425, upper whisker=0.0506}] coordinates {(7,0.1152)};
\addplot+[draw=figred, fill=figred!28, boxplot prepared={draw position=8, lower whisker=0.0292, lower quartile=3.6489, median=14.9728, upper quartile=21.9369, upper whisker=27.863}] coordinates {};
\addplot+[boxplot prepared={draw position=9, lower whisker=0, lower quartile=0.0035, median=0.0093, upper quartile=0.018, upper whisker=0.0377}] coordinates {(9,0.1087)};
\addplot+[boxplot prepared={draw position=10, lower whisker=0, lower quartile=0, median=0, upper quartile=0.0043, upper whisker=0.0057}] coordinates {(10,0.0375) (10,0.0864)};
\addplot+[boxplot prepared={draw position=11, lower whisker=0, lower quartile=0.0397, median=0.4396, upper quartile=0.5239, upper whisker=0.5761}] coordinates {};
\addplot+[boxplot prepared={draw position=12, lower whisker=0, lower quartile=0.0002, median=0.0096, upper quartile=0.0356, upper whisker=0.0564}] coordinates {};
\addplot+[boxplot prepared={draw position=13, lower whisker=0.0026, lower quartile=0.0287, median=0.0597, upper quartile=0.0877, upper whisker=0.0894}] coordinates {(13,0.3007) (13,0.5657)};
\addplot+[boxplot prepared={draw position=14, lower whisker=0, lower quartile=0, median=0, upper quartile=0.0043, upper whisker=0.0063}] coordinates {(14,0.0229)};
\addplot+[draw=figred, fill=figred!28, boxplot prepared={draw position=15, lower whisker=3.5928, lower quartile=4.4168, median=5.7875, upper quartile=6.4326, upper whisker=7.0725}] coordinates {(15,0)};
\addplot+[boxplot prepared={draw position=16, lower whisker=0.0042, lower quartile=0.0448, median=0.0709, upper quartile=0.0912, upper whisker=0.1604}] coordinates {};
\addplot+[boxplot prepared={draw position=17, lower whisker=0.1128, lower quartile=0.1868, median=0.2319, upper quartile=0.2677, upper whisker=0.2773}] coordinates {(17,0.5584) (17,0.562)};
\addplot+[boxplot prepared={draw position=18, lower whisker=0.2057, lower quartile=0.2586, median=0.3519, upper quartile=0.4578, upper whisker=0.5187}] coordinates {};
\addplot+[boxplot prepared={draw position=19, lower whisker=0, lower quartile=0.0098, median=0.0565, upper quartile=0.0836, upper whisker=0.178}] coordinates {(19,0.2381)};
\addplot+[draw=figred, fill=figred!28, boxplot prepared={draw position=20, lower whisker=0, lower quartile=0.0187, median=2.3076, upper quartile=7.5352, upper whisker=16.8274}] coordinates {};
\addplot+[boxplot prepared={draw position=21, lower whisker=0, lower quartile=0, median=0, upper quartile=0, upper whisker=0}] coordinates {};
\addplot+[boxplot prepared={draw position=22, lower whisker=0, lower quartile=0.0165, median=0.0482, upper quartile=0.2456, upper whisker=0.432}] coordinates {};
\addplot+[boxplot prepared={draw position=23, lower whisker=0, lower quartile=0, median=0, upper quartile=0, upper whisker=0}] coordinates {(23,0.0443) (23,0.1043)};
\addplot+[boxplot prepared={draw position=24, lower whisker=0, lower quartile=0, median=0, upper quartile=0, upper whisker=0}] coordinates {};
\addplot+[boxplot prepared={draw position=25, lower whisker=0, lower quartile=0.0023, median=0.1212, upper quartile=0.2158, upper whisker=0.2166}] coordinates {(25,0.8619) (25,0.7297)};
\addplot+[boxplot prepared={draw position=26, lower whisker=0.7467, lower quartile=1.3059, median=1.4327, upper quartile=1.7503, upper whisker=2.1572}] coordinates {(26,0.2074)};
\addplot+[boxplot prepared={draw position=27, lower whisker=0, lower quartile=0, median=0, upper quartile=0.0063, upper whisker=0.0085}] coordinates {(27,0.0504) (27,0.1074)};
\addplot+[draw=figred, fill=figred!28, boxplot prepared={draw position=28, lower whisker=5.9044, lower quartile=6.2674, median=6.7222, upper quartile=7.3817, upper whisker=7.6137}] coordinates {};
\addplot+[boxplot prepared={draw position=29, lower whisker=0, lower quartile=0, median=0, upper quartile=0, upper whisker=0}] coordinates {};
\addplot+[boxplot prepared={draw position=30, lower whisker=0, lower quartile=0, median=0, upper quartile=0.0196, upper whisker=0.0262}] coordinates {(30,0.836) (30,0.0516)};
\addplot+[boxplot prepared={draw position=31, lower whisker=0, lower quartile=0, median=0.0801, upper quartile=0.3954, upper whisker=0.5401}] coordinates {};
\addplot+[boxplot prepared={draw position=32, lower whisker=0.0097, lower quartile=0.0142, median=0.0206, upper quartile=0.0228, upper whisker=0.0229}] coordinates {(32,0.0656) (32,0.0011) (32,0.0411)};
\addplot+[boxplot prepared={draw position=33, lower whisker=0, lower quartile=0.0246, median=0.147, upper quartile=0.1922, upper whisker=0.4408}] coordinates {};
\addplot+[boxplot prepared={draw position=34, lower whisker=0, lower quartile=0, median=0, upper quartile=0.7922, upper whisker=1.1767}] coordinates {};
\addplot+[boxplot prepared={draw position=35, lower whisker=0, lower quartile=0, median=0, upper quartile=0, upper whisker=0}] coordinates {};
\addplot+[boxplot prepared={draw position=36, lower whisker=0.6186, lower quartile=0.8203, median=1.1753, upper quartile=1.2317, upper whisker=1.3372}] coordinates {};
\addplot+[boxplot prepared={draw position=37, lower whisker=0, lower quartile=0, median=0.0001, upper quartile=0.0023, upper whisker=0.003}] coordinates {(37,0.007)};
\addplot+[boxplot prepared={draw position=38, lower whisker=0, lower quartile=0, median=0.0278, upper quartile=0.2477, upper whisker=0.5963}] coordinates {};
\addplot+[draw=figred, fill=figred!28, boxplot prepared={draw position=39, lower whisker=0, lower quartile=0.4137, median=0.7014, upper quartile=1.1567, upper whisker=1.281}] coordinates {(39,3.9432)};
\addplot+[boxplot prepared={draw position=40, lower whisker=0.0494, lower quartile=0.1184, median=0.2172, upper quartile=0.2449, upper whisker=0.3776}] coordinates {};
\addplot+[draw=figred, fill=figred!28, boxplot prepared={draw position=41, lower whisker=0.05, lower quartile=0.1993, median=1.1401, upper quartile=1.7726, upper whisker=3.3368}] coordinates {};
\addplot+[boxplot prepared={draw position=42, lower whisker=0, lower quartile=0.0043, median=0.0852, upper quartile=0.1209, upper whisker=0.1743}] coordinates {};
\addplot+[boxplot prepared={draw position=43, lower whisker=1.6401, lower quartile=1.7037, median=1.7258, upper quartile=1.7509, upper whisker=1.7861}] coordinates {};
\addplot+[boxplot prepared={draw position=44, lower whisker=0.1492, lower quartile=0.1517, median=0.1762, upper quartile=0.1832, upper whisker=0.196}] coordinates {(44,0.2383)};
\addplot+[boxplot prepared={draw position=45, lower whisker=0, lower quartile=0, median=0, upper quartile=0, upper whisker=0}] coordinates {};
\addplot+[boxplot prepared={draw position=46, lower whisker=0, lower quartile=0, median=0, upper quartile=0, upper whisker=0}] coordinates {(46,0.0104) (46,0.0007)};
\addplot+[draw=figred, fill=figred!28, boxplot prepared={draw position=47, lower whisker=2.3223, lower quartile=3.5557, median=4.2277, upper quartile=4.8038, upper whisker=6.6401}] coordinates {(47,7.8159)};
\addplot+[boxplot prepared={draw position=48, lower whisker=0, lower quartile=0.0188, median=0.2355, upper quartile=1.6004, upper whisker=1.6578}] coordinates {};
\addplot+[boxplot prepared={draw position=49, lower whisker=0.0127, lower quartile=0.0802, median=0.1013, upper quartile=0.2005, upper whisker=0.2993}] coordinates {(49,0.4324)};
\end{axis}
\end{tikzpicture}
\caption{Per-benchmark IPC loss of the BGSP, Gaze/Entangling/Mockingjay, across ten 20M-instruction phases. Red boxes contain at least one phase above the dashed 2.5\% line.}
\label{fig:loss_dist_single}
\end{figure}

\emph{Application-specific analysis.} The aggregate view shows that the tail exists; the per-application view shows that it is not an artifact of one benchmark. Figure~\ref{fig:loss_dist_single} plots the BGSP loss to oracle separately for all 49 applications, with one box plot per application and ten phase samples per box. Some applications stay close to zero across all phases, while others show consistently high loss or isolated outliers. Eight benchmarks contribute at least one phase above 2.5\%, so the tail spans multiple workloads rather than being driven by a single unusual trace. It includes applications with high median loss, applications with substantial within-trace variation, and applications with only one or two costly phases.

\emph{Comparison to IP-stride/LRU.} Before further analysis, we place the modern policies in context. Relative to the IP-stride/LRU reference point, the modern-policy oracle improves mean IPC by 30.09\%, exceeds 2.5\% gain in 470 phases, and averages 31.31\% gain across those phases. Modern prefetching and replacement therefore capture the dominant, broadly available improvement over the reference system. The average 1--2\% composition gap is the remaining opportunity among strong candidates, and the per-application tail explains why that smaller gap is still worth understanding.

\subsection{Does One Static Configuration Dominate?}

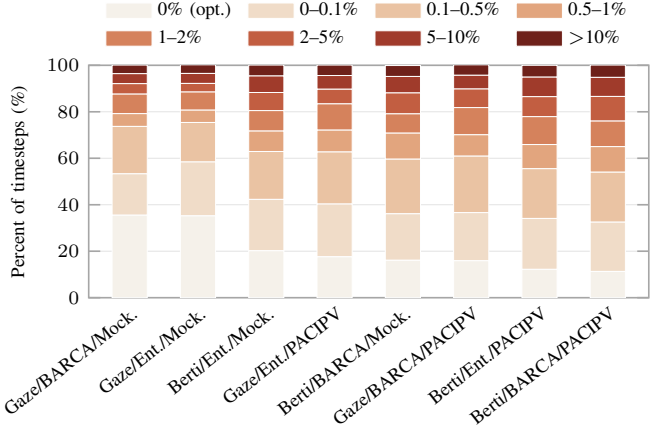
\begin{figure}[tb]
\centering
\resizebox{\columnwidth}{!}{
\begin{tikzpicture}
\begin{axis}[
  hpcafig,
  width=\columnwidth, height=4.6cm,
  ybar stacked, bar width=13pt,
  ymin=0, ymax=100,
  ytick={0,20,40,60,80,100},
  ylabel={Percent of timesteps (\%)},
  xmin=0.4, xmax=8.6,
  xtick={1,...,8},
  xticklabels={Gaze/BARCA/Mock.,Gaze/Ent./Mock.,Berti/Ent./Mock.,Gaze/Ent./PACIPV,
               Berti/BARCA/Mock.,Gaze/BARCA/PACIPV,Berti/Ent./PACIPV,Berti/BARCA/PACIPV},
  xticklabel style={font=\scriptsize, rotate=38, anchor=north east, yshift=1pt, xshift=2pt},
  legend style={
    at={(0.5,1.03)}, anchor=south,
    legend columns=4,
    /tikz/every even column/.append style={column sep=4pt},
    font=\scriptsize, draw=none, fill=none,
  },
  legend image code/.code={\draw[#1, draw=gray!80, line width=0.3pt] (0cm,-0.06cm) rectangle (0.22cm,0.10cm);},
  area legend,
]
\addplot[ybar, fill=sev1, draw=white, line width=0.3pt] coordinates {(1,35.5)(2,35.2)(3,20.2)(4,17.6)(5,16.1)(6,15.9)(7,12.2)(8,11.3)};
\addplot[ybar, fill=sev2, draw=white, line width=0.3pt] coordinates {(1,17.8)(2,23.2)(3,22.1)(4,22.7)(5,20.0)(6,20.7)(7,21.9)(8,21.2)};
\addplot[ybar, fill=sev3, draw=white, line width=0.3pt] coordinates {(1,20.4)(2,17.0)(3,20.6)(4,22.4)(5,23.5)(6,24.3)(7,21.4)(8,21.5)};
\addplot[ybar, fill=sev4, draw=white, line width=0.3pt] coordinates {(1,5.5)(2,5.3)(3,8.8)(4,9.4)(5,11.2)(6,9.2)(7,10.4)(8,11.0)};
\addplot[ybar, fill=sev5, draw=white, line width=0.3pt] coordinates {(1,8.4)(2,7.8)(3,8.8)(4,11.3)(5,8.3)(6,11.7)(7,12.0)(8,11.0)};
\addplot[ybar, fill=sev6, draw=white, line width=0.3pt] coordinates {(1,4.5)(2,3.7)(3,7.8)(4,6.3)(5,9.0)(6,8.0)(7,8.6)(8,10.6)};
\addplot[ybar, fill=sev7, draw=white, line width=0.3pt] coordinates {(1,4.3)(2,4.3)(3,7.1)(4,5.9)(5,7.1)(6,5.9)(7,8.4)(8,8.2)};
\addplot[ybar, fill=sev8, draw=white, line width=0.3pt] coordinates {(1,3.7)(2,3.7)(3,4.7)(4,4.5)(5,4.7)(6,4.5)(7,5.1)(8,5.3)};
\legend{0\% (opt.),0--0.1\%,0.1--0.5\%,0.5--1\%,1--2\%,2--5\%,5--10\%,$>$10\%}
\end{axis}
\end{tikzpicture}}
\caption{IPC-loss ranges for all eight static configurations. No configuration is uniformly best; each leaves a distinct mix of near-equal and costly phases. Ent.\ denotes Entangling and Mock.\ denotes Mockingjay.}
\label{fig:all_policies}
\end{figure}

The previous analysis focuses on the BGSP, which naturally raises the question of whether the simplest answer is to build that configuration and stop. The answer is no. Figure~\ref{fig:all_policies} compares all eight configurations against the same phase oracle. Every configuration loses to an alternative in some phases, and every configuration reaches the oracle in a nonzero fraction of phases. A dominant configuration would eliminate oracle selections of the alternatives and place nearly all of its own phases in the lowest-loss range. Instead, all eight show a mixture of exact wins, near-equal outcomes, and larger losses.

The frequency and severity of the losses differ, but no fixed point eliminates the opportunity. This distinction matters because a low average loss does not mean the policies are interchangeable. Most configuration choices have limited IPC impact, while a smaller set of phases accounts for most recoverable performance. Runtime selection should therefore be judged by IPC loss, not just by how often it agrees with an oracle label.

\subsection{Ranking the Policies}

\begin{figure}[tb]
\centering
\resizebox{\columnwidth}{!}{
\begin{tikzpicture}
\begin{axis}[
  hpcafig,
  width=0.92\columnwidth, height=4.4cm,
  xmin=0, xmax=39,
  xtick={0,10,20,30},
  xlabel={Percent of timesteps optimal (\%)},
  xlabel style={yshift=2pt},
  xmajorgrids, ymajorgrids=false,
  grid style={gray!22, very thin},
  symbolic y coords={Berti/BARCA/PACIPV,Gaze/BARCA/PACIPV,Berti/BARCA/Mock.,
                     Berti/Ent./PACIPV,Gaze/Ent./PACIPV,Gaze/BARCA/Mock.,
                     Berti/Ent./Mock.,Gaze/Ent./Mock.},
  ytick={Berti/BARCA/PACIPV,Gaze/BARCA/PACIPV,Berti/BARCA/Mock.,Berti/Ent./PACIPV,Gaze/Ent./PACIPV,Gaze/BARCA/Mock.,Berti/Ent./Mock.,Gaze/Ent./Mock.},
  enlarge y limits=0.09,
  nodes near coords={\pgfmathprintnumber[fixed,precision=1,zerofill]{\pgfplotspointmeta}},
  every node near coord/.append style={font=\scriptsize, anchor=west, xshift=1pt},
  legend style={at={(0.97,0.05)}, anchor=south east},
  legend image code/.code={\draw[#1, line width=0.3pt] (0cm,-0.06cm) rectangle (0.22cm,0.10cm);},
  clip=false,
]
\addplot[xbar, bar shift=0pt, bar width=8pt, point meta=rawx, fill=figblue!75, draw=figblue, line width=0.35pt]
  coordinates {(33.5,Gaze/Ent./Mock.)(15.3,Gaze/BARCA/Mock.)(11.2,Gaze/Ent./PACIPV)(3.7,Gaze/BARCA/PACIPV)};
\addplot[xbar, bar shift=0pt, bar width=8pt, point meta=rawx, fill=gray!40, draw=figgray, line width=0.35pt]
  coordinates {(19.2,Berti/Ent./Mock.)(9.0,Berti/Ent./PACIPV)(5.1,Berti/BARCA/Mock.)(3.1,Berti/BARCA/PACIPV)};
\legend{Gaze L1D,Berti L1D}
\end{axis}
\end{tikzpicture}}
\caption{Oracle win frequency for each complete configuration, grouped by L1D prefetcher family. The two most frequent choices differ only between Gaze and Berti.}
\label{fig:policy_freq}
\end{figure}
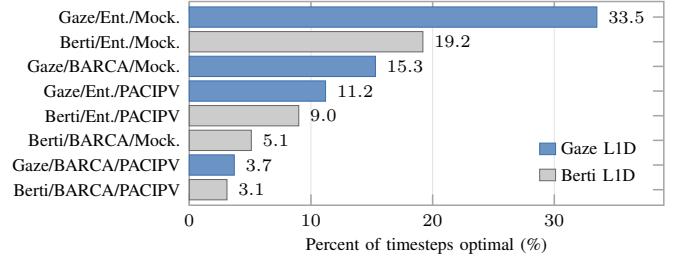

Oracle choices reveal where the eight-way space is redundant. Figure~\ref{fig:policy_freq} confirms that the BGSP, Gaze/Entangling/Mockingjay, wins 33.47\% of phases. Berti/Entangling/Mockingjay wins 19.18\%. Together, those two configurations account for 52.65\% of oracle decisions, and they differ only in the L1D prefetcher. Adding the corresponding BARCA choices raises the covered share to about 73.1\% of phases. \footnote{Figure~\ref{fig:all_policies} computes percentage of optimal timesteps differently from Figure~\ref{fig:policy_freq}; the former is a percentage of raw count while the latter looks at the percentage of distinct timesteps for which a particular configuration is optimal.}

Frequency alone is not enough to choose a runtime action set. A rarely selected configuration can still matter if it wins a small number of high-margin phases, and a frequent configuration can add little if most of its wins are ties or near ties. The next step is therefore not to pick the two tallest bars, but to ask which small subset preserves oracle performance, creates useful switching opportunities, and keeps the hardware interface narrow.

\subsection{Selecting a Two-Configuration Subset}
\label{sec:compression}

Selecting among all eight configurations would be expensive for a modest average opportunity. It would require multiple prefetchers and replacement policies, arbitration among their requests, several sets of policy-local state, and validation of every supported interaction. The observation problem grows with the candidate set as well: the core sees the IPC and cache state produced by the configuration it actually executed, but not the seven inactive outcomes. A practical design should therefore ask whether a smaller subset captures most of the oracle benefit before it designs a selector. For this paper, we simplify the question to 'what if we have just two policies to choose from?'

The two highest-ranked static configurations are not necessarily the best pair. We rank pairs using three criteria. \emph{Oracle coverage} measures the loss when the full eight-way maximum is restricted to the pair. \emph{Static-member gain} measures how much perfect switching improves over the better member of the pair. \emph{Complementarity} requires both candidates to win meaningful intervals rather than differ mostly through equal-IPC outcomes.

Simply picking policies that rank close to each other according to the mean IPC loss to oracle, fails these metrics. Several high-ranking configurations belong to the same Berti-family region of the design space and differ only by L1I prefetching or L2 replacement. Their mean IPC ranks are different, but their pairwise gaps are below 0.1\%, so perfect selection between them would provide little additional benefit.

Based on these criteria, our selected policy pair fixes BARCA (L1I prefetching) and PACIPV (L2 replacement) and changes only the L1D prefetcher: Gaze/BARCA/PACIPV and Berti/BARCA/PACIPV. This choice bounds state management to two L1D engines while instruction prefetching and L2 replacement continue uninterrupted. On the 4,900 200K-instruction windows used for selector replay, Gaze wins 51.55\% of windows, Berti wins 44.71\%, and the policies tie in 3.73\%. Gaze is 2.74\% faster when it wins; when Berti wins, selecting Gaze incurs 3.41\% average loss. The less frequent Berti outcome therefore carries the larger conditional penalty, which makes the pair useful for selection rather than merely balanced by count. Subsequent sections evaluate how much of this binary Berti/Gaze opportunity can be recovered from signals available to a runtime selector. The opportunity is highly compressible: a Berti/Gaze pair that changes only the L1D prefetcher comes within 0.039\% aggregate IPC of the eight-configuration oracle across all 4,900 windows.

\subsection{Caveats and Positioning}
This paper intentionally narrows the runtime selector to two configurations, but the methodology is not limited to binary selectors. The broader workflow is to measure the full composition space, then choose a candidate subset using oracle coverage, gain over the better static member, complementarity, and switching scope. For this policy set, those criteria identify a binary L1D-prefetcher choice that preserves the useful opportunity while keeping state, arbitration, and validation costs bounded. A different policy set or workload mix could justify more candidates; the remaining sections evaluate the narrow target exposed by this analysis rather than claiming that two choices are always sufficient.


\section{Runtime Selector Design Space}
\label{sec:selector_design}

\subsection{Binary Selection Interface}

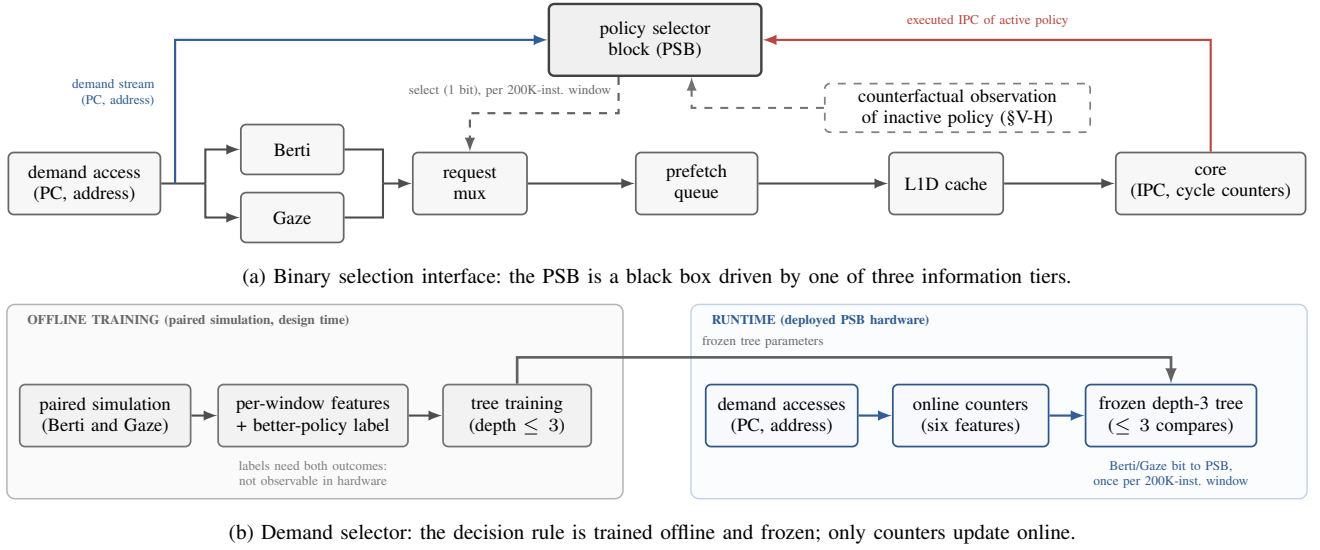
\begin{figure*}[tb]
\centering
\begin{tikzpicture}[
  font=\scriptsize,
  block/.style={
    draw=black!60,
    fill=black!3,
    rounded corners=2pt,
    align=center,
    minimum height=0.82cm,
    inner xsep=3pt,
    inner ysep=2pt,
    line width=0.65pt
  },
  psb/.style={
    block,
    draw=black!75,
    fill=black!6,
    line width=0.9pt
  },
  hypo/.style={
    block,
    dashed,
    draw=black!50,
    fill=white
  },
  dataarrow/.style={-latex, draw=black!70, line width=0.78pt},
  demandarrow/.style={-latex, draw=figblue!85!black, line width=0.82pt},
  rewardarrow/.style={-latex, draw=figred!85!black, line width=0.82pt},
  hypoarrow/.style={-latex, dashed, draw=black!55, line width=0.78pt},
  controlarrow/.style={-latex, dashed, draw=black!62, line width=0.82pt}
]

\node[block, text width=1.80cm] (access)
  at (-7.55,0) {demand access\\(PC, address)};
\node[block, text width=1.15cm, minimum height=0.66cm] (berti)
  at (-4.80,0.45) {Berti};
\node[block, text width=1.15cm, minimum height=0.66cm] (gaze)
  at (-4.80,-0.45) {Gaze};
\node[block, text width=1.30cm] (mux)
  at (-2.45,0) {request\\mux};
\node[block, text width=1.40cm] (queue)
  at (0.55,0) {prefetch queue};
\node[block, text width=1.30cm] (cache)
  at (3.85,0) {L1D cache};
\node[block, text width=2.30cm] (core)
  at (7.35,0) {core\\(IPC, cycle counters)};
\coordinate (tap) at (-6.35,0);
\coordinate (fork) at (-5.95,0);
\draw[draw=black!70, line width=0.78pt] (access.east) -- (fork);
\draw[dataarrow] (fork) |- (berti.west);
\draw[dataarrow] (fork) |- (gaze.west);
\coordinate (merge) at (-3.60,0);
\draw[draw=black!70, line width=0.78pt] (berti.east) -| (merge);
\draw[draw=black!70, line width=0.78pt] (gaze.east) -| (merge);
\draw[dataarrow] (merge) -- (mux.west);
\draw[dataarrow] (mux.east) -- (queue.west);
\draw[dataarrow] (queue.east) -- (cache.west);
\draw[dataarrow] (cache.east) -- (core.west);

\node[psb, text width=2.60cm, minimum height=0.95cm] (psb)
  at (0.00,1.90) {policy selector\\block (PSB)};

\draw[demandarrow] (tap) -- (-6.35,1.90) -- (psb.west);
\node[font=\tiny, text=figblue!82!black, anchor=east, align=right]
  at (-6.50,1.20) {demand stream\\(PC, address)};

\draw[rewardarrow] (core.north) -- (7.35,1.90) -- (psb.east);
\node[font=\tiny, text=figred!82!black, anchor=south]
  at (4.40,1.95) {executed IPC of active policy};


\node[hypo, text width=3.30cm, minimum height=0.62cm] (cf)
  at (4.00,1.00) {counterfactual observation\\of inactive policy (\S\ref{sec:ipu})};
\coordinate (psb-bottom-twothirds)
  at ($(psb.south west)!0.67!(psb.south east)$);
\draw[hypoarrow]
  (cf.west)
  -- (psb-bottom-twothirds |- cf.west)
  -- (psb-bottom-twothirds);

\coordinate (psb-bottom-third)
  at ($(psb.south west)!0.33!(psb.south east)$);
\draw[controlarrow]
  (psb-bottom-third) -- (psb-bottom-third |-,0.93) -| (mux.north);
\node[font=\tiny, text=black!58, anchor=south]
  at (-1.92,1.02) {select (1 bit), per 200K-inst.\ window};  

\end{tikzpicture}

\par\smallskip
{\footnotesize (a) Binary selection interface: the PSB is a black box driven by one of three information tiers.}
\par\medskip

\begin{tikzpicture}[
  font=\scriptsize,
  block/.style={
    draw=black!60,
    fill=black!3,
    rounded corners=2pt,
    align=center,
    minimum height=0.82cm,
    inner xsep=3pt,
    inner ysep=2pt,
    line width=0.65pt
  },
  offline/.style={
    block,
    draw=black!55,
    fill=black!5
  },
  runtime/.style={
    block,
    draw=figblue!78!black,
    fill=figblue!4
  },
  dataarrow/.style={-latex, draw=black!70, line width=0.78pt},
  demandarrow/.style={-latex, draw=figblue!85!black, line width=0.82pt},
  shiparrow/.style={-latex, draw=black!62, line width=1.0pt}
]

\draw[rounded corners=3pt, draw=black!30, fill=black!2, line width=0.55pt]
  (-8.60,-0.95) rectangle (-0.45,1.62);
\draw[rounded corners=3pt, draw=figblue!30, fill=figblue!2, line width=0.55pt]
  (0.45,-0.95) rectangle (8.60,1.62);
\node[font=\tiny\bfseries, text=black!60, anchor=west]
  at (-8.45,1.40) {OFFLINE TRAINING (paired simulation, design time)};
\node[font=\tiny\bfseries, text=figblue!82!black, anchor=west]
  at (0.60,1.40) {RUNTIME (deployed PSB hardware)};

\node[offline, text width=2.05cm] (sim)
  at (-7.30,0.15) {paired simulation\\(Berti and Gaze)};
\node[offline, text width=2.30cm] (label)
  at (-4.55,0.15) {per-window features\\+ better-policy label};
\node[offline, text width=1.75cm] (train)
  at (-1.85,0.15) {tree training\\(depth $\le 3$)};
\draw[dataarrow] (sim.east) -- (label.west);
\draw[dataarrow] (label.east) -- (train.west);
\node[font=\tiny, text=black!50, anchor=north, align=center]
  at (-4.55,-0.32) {labels need both outcomes:\\not observable in hardware};

\node[runtime, text width=1.80cm] (dem)
  at (1.65,0.15) {demand accesses\\(PC, address)};
\node[runtime, text width=1.85cm] (cnt)
  at (4.15,0.15) {online counters\\(six features)};
\node[runtime, text width=2.05cm] (tree)
  at (6.80,0.15) {frozen depth-3 tree\\($\le 3$ compares)};
\draw[demandarrow] (dem.east) -- (cnt.west);
\draw[demandarrow] (cnt.east) -- (tree.west);
\node[font=\tiny, text=figblue!82!black, anchor=north, align=center]
  at (6.80,-0.32) {Berti/Gaze bit to PSB,\\once per 200K-inst.\ window};

\draw[shiparrow] (train.north) -- (-1.85,0.92) -- (6.80,0.92) -- (tree.north);
\node[font=\tiny, text=black!58, rounded corners=1.5pt,
      inner xsep=3pt, inner ysep=1.8pt]
  at (1.40,1.12) {frozen tree parameters};
\end{tikzpicture}

\par\smallskip
{\footnotesize (b) Demand selector: the decision rule is trained offline and frozen; only counters update online.}
\caption{Policy selector block. (a) Hardware interface: the PSB is a black box that emits one Berti/Gaze select bit per window; demand-stream monitoring, executed-performance feedback, and (hypothetical) counterfactual observation are alternative information tiers that could drive it, not coexisting circuits. (b) Demand-selector instantiation: the decision tree is trained offline from paired simulation and deployed frozen; at runtime only the feature counters update, and the tree is traversed once per window.}
\label{fig:selector_hardware}
\end{figure*}

The candidate reduction in Section~\ref{sec:compression} leaves two complete configurations that differ only in the L1D prefetcher. So, the hardware interface is narrow: both Berti and Gaze are present, but a one-bit select signal decides which engine can issue requests into the shared prefetch queue and L1D cache. We call the hardware that produces this bit the policy selector block (PSB). Instruction prefetching, L2 replacement, the cache hierarchy, and the memory system remain unchanged. Figure~\ref{fig:selector_hardware} shows this interface. Regardless of how the PSB computes the bit, the memory system sees the same request mux and one active L1D prefetcher at a time.

The inactive prefetcher still raises a state-management question. Its tables can be retained and frozen, allowing a faster restart but consuming storage and possibly preserving stale state. They can instead be reset, which simplifies implementation but forces the prefetcher to relearn after a switch. A third option is to update a limited amount of passive state while preventing the inactive engine from issuing requests. We do not choose among these policies here. Section~\ref{sec:evaluation} will measure switching frequency and minimum residence time, which bound how often any chosen state-management policy would pay its cost.

\subsection{Runtime Information Sources}
The central design question is what the PSB is allowed to see before it chooses a prefetcher. We organize selectors into three information tiers: executed-performance feedback, demand-stream monitoring, and counterfactual observation.

\paragraph{Executed-performance feedback.}
Reward feedback uses the least policy-specific signal: IPC from the prefetcher currently in control. We evaluate a small Upper Confidence Bound (UCB) rule, a standard exploration/exploitation algorithm from reinforcement learning. Informally, the UCB policy does the following: pick the recent best policy, but occasionally retry the other policy to check whether it has become better. The PSB keeps the last few IPC observations for each prefetcher and adds a small retry bonus to the less recently tested one. The hardware inputs are simple because instruction and cycle counters already exist. The cost is not the arithmetic in the UCB rule; the cost is that refreshing an inactive-policy estimate requires giving that policy real control of the L1D prefetch stream for a window, and that sampled window may run slower.

\paragraph{Demand-stream monitoring.}
In the second tier, the PSB contains a passive demand monitor that observes PCs and demand addresses before prefetching changes cache state. This makes the signal action independent: the monitor sees the same stream whether Berti or Gaze is currently allowed to issue. We evaluate two ways to turn that stream into a decision. The compact approach compares policy-oriented confidence counters, one summarizing Berti-like per-PC delta regularity and one summarizing Gaze-like footprint recurrence. The richer approach keeps address-derived counters grouped by purpose: some measure the local delta patterns Berti exploits, some measure the recurring spatial footprints Gaze exploits, and some measure how interleaved or concentrated the access stream is across PCs and regions.

\begin{table*}[tb]
\centering
\caption{Demand-flow Feature Vector Components.}
\label{tab:features}
\begin{tabular}{lcc}
\toprule
Name & Description \\
\midrule
\textit{sample\_refs} & Demand memory references  \\
\textit{unique\_lines} & Different cache lines accessed  \\
\textit{delta\_obs} & Deltas generated by same-PC consecutive visits\\
\textit{avg\_region\_density} & Average proportion of cache lines accessed per memory region (i.e. a 4 KB address block)  \\
\textit{avg\_pcs\_per\_region} & Average number of different demand PCs corresponding to each memory region \\
\textit{footprint\_recurrence\_rate} & Proportion of regions where the footprint reappears\\
\bottomrule
\end{tabular}
\end{table*}

A deployed demand selector can be implemented as a small counter block plus a frozen decision tree. The counters are filled online as the program runs, similar in spirit to branch history tables that accumulate local execution history. The difference is that the rule interpreting those counters is fixed before deployment. During offline training, each training window is converted into a six-scalar demand-flow feature vector (shown in Table~\ref{tab:features}) and labeled with the better of Berti and Gaze from paired simulation. The processor cannot dynamically produce that label unless it somehow observes both outcomes, so the labels come from simulation on previously seen benchmarks, and Section~\ref{sec:evaluation} tests transfer on held-out benchmarks. PACIPV follows a similar offline-training pattern, but for a different feature set and prediction target. At runtime, the monitor materializes the same six feature values after the sampling prefix and traverses the frozen tree. Depth three, as used here, is a maximum root-to-leaf depth: the unpruned tree has 15 entries, with seven internal compare nodes and eight leaf nodes. Each internal entry stores a feature ID, threshold, and left/right child pointers; each leaf stores the policy bit, Berti or Gaze. Inference therefore performs at most three comparisons of one selected feature value against a stored threshold, then returns the leaf policy once per 200K-instruction window rather than on the per-access path. The monitor does not generate prefetch addresses, allocate cache lines, consume MSHRs, or use memory bandwidth. It only watches demand accesses and produces the same one-bit Berti/Gaze decision used by the interface in Figure~\ref{fig:selector_hardware}.

\paragraph{Counterfactual observation.}
Counterfactual information asks for the signal the previous two tiers lack: what would the inactive prefetcher have done if it had been active? That includes candidate requests, usefulness, timeliness, queue pressure, possible cache displacements, etc. An IPU-style substrate~\cite{mcdougall2025ipu} - which has already been shown to be able to emulate the behavior of the entangled prefetcher - or a shadow monitor could expose some of this information while preventing the inactive engine from perturbing the real hierarchy. Section~\ref{sec:ipu} treats this as an information upper bound rather than a practical design in this section: it starts with an exact winner signal, then adds delay, error, and sparse observation to determine how good a real counterfactual monitor would need to be. We leave to future work whether a specialized, practical, and very small counterfactual estimator is feasible.

\section{Practical Selector Evaluation}
\label{sec:evaluation}

Section~\ref{sec:selector_design} defines three ways a policy selector block (PSB) can obtain information before choosing Berti or Gaze. This evaluation asks how much of the two-policy oracle opportunity each information source can recover, and where the practical boundary lies before adding hardware that observes the inactive prefetcher. We proceed from the cheapest signal to the strongest bound: chronological chosen-policy feedback, demand-stream monitoring through compact confidence values and then a six-feature depth-three tree, sensitivity studies for prefix length, feature set, residence time, and policy-pair choice, and finally an ideal counterfactual winner signal.

\subsection{Simulation Setup and Metrics}

All results use the same ChampSim timing model and 4,900 200K-instruction windows introduced in Section~\ref{sec:compression}. Conceptually, the PSB is invoked at window boundaries, chooses Berti or Gaze, and ChampSim provides the IPC of the policy that was active for that interval. This organization lets us apply several selector rules to the same ChampSim windows and compare them with static and oracle references; it is not a separate simulator or a new timing model. The executed-performance selector is evaluated chronologically over all 49 traces because it learns online from the IPC of the policy it actually ran. Learned demand-stream selectors use a benchmark-disjoint test: all windows from 39 benchmarks train the tree, and all windows from the remaining 10 benchmarks are held out for evaluation. Splitting by benchmark, rather than randomly mixing windows, prevents the tree from seeing one part of a program during training and another part of the same program during testing. The primary demand model uses all six demand-flow features, maximum depth three, and a deterministic training seed.

For each window, losses are computed from the ChampSim IPC of the selected policy and measured against the pairwise Berti/Gaze oracle for that same interval (BARCA and PACIPV remain fixed in the pairwise oracle configuration). Accuracy is the fraction of decisions that reach the pairwise-oracle IPC; if Berti and Gaze tie, either selection is counted correct. Accuracy is useful for debugging a selector, but IPC loss is the primary metric because a wrong choice with a 0.01\% margin and a wrong choice with a 10\% margin should not count the same. We also report recovered oracle gap: of the loss left by a fixed static choice, what fraction does the selector remove?
\begin{equation}
Recovery =
\frac{Loss_{\mathit{static}}-Loss_{\mathit{selector}}}
     {Loss_{\mathit{static}}}\times 100\%.
\end{equation}
We always use the Gaze/BARCA/PACIPV policy configuration as our static baseline, which we dub \textit{always-Gaze}. Always-Gaze is computed over the same population as the selector in that table. Thus, always-Gaze has 1.53\% loss over all 4,900 windows in the UCB experiment and 1.237\% loss over the 1,000 windows across the 10 held-out benchmarks in the demand-stream experiments. In addition to loss, we compute the selector's aggregate gain over the always-Gaze policy:

\begin{equation}
Gain_{\mathrm{agg}}(s,b)= \left( \frac{\sum_{w\in W} IPC(s,w)} {\sum_{w\in W} IPC(b,w)}-1 \right)\times100\%
\end{equation}
where \(b\) is always-Gaze and both results use the same window set \(W\). The demand tree never receives inactive-policy IPC, usefulness, timeliness, pollution, bandwidth, MSHR pressure, or cache outcomes as inputs. 

\subsection{Executed-Performance Feedback}

The cheapest PSB watches the IPC it is already producing. We instantiate this tier with sliding-window Upper Confidence Bound (UCB): keep a short recent IPC history for each candidate, pick the policy with the larger recent score, and occasionally retry the other policy to check whether the winner has changed. The reported point keeps the three most recent rewards per candidate and uses exploration coefficient $c=0.03$. We also sweep history lengths of 3, 5, 10, and 20 windows and $c$ from 0.01 to 0.20. A shorter history reacts sooner after winner changes, while a longer history smooths IPC noise; a larger $c$ refreshes inactive estimates more often, but spends more windows deliberately sampling a policy that may be worse.

\begin{figure*}[tb]
\centering
\includegraphics[width=0.9\textwidth]{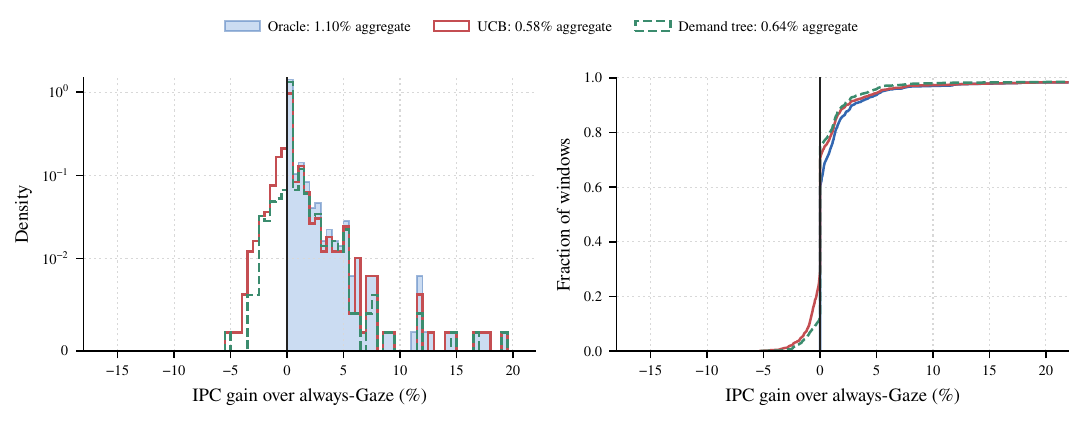}
\caption{Distribution of IPC gain over the pair-static always-Gaze baseline on the 1,000 held-out windows. The two-policy oracle, sliding-window UCB, and address tree gain 1.10\%, 0.58\%, and 0.64\% aggregate IPC, respectively.}
\label{fig:selector_gaze_distribution}
\end{figure*}

Over all 4,900 windows, UCB lowers mean loss from 1.53\% for the pair-static always-Gaze baseline to 0.407\%, a $3.75\times$ reduction, and reaches 68.41\% accuracy. Therefore, UCB recovers 73.4\% of the pairwise-oracle gap left by the static choice. Its aggregate IPC gain over always-Gaze is 1.25\%. Figure~\ref{fig:selector_gaze_distribution} compares the two-policy oracle, UCB, and the address tree against always-Gaze on the common 1,000-window held-out population used by the learned selector. The three methods gain 1.10\%, 0.58\%, and 0.64\% aggregate IPC, respectively. The distributions show that both practical selectors capture much of the positive oracle mass, while their left tails expose windows in which stale reward estimates or an incorrect demand-based prediction select the lower-performing policy.

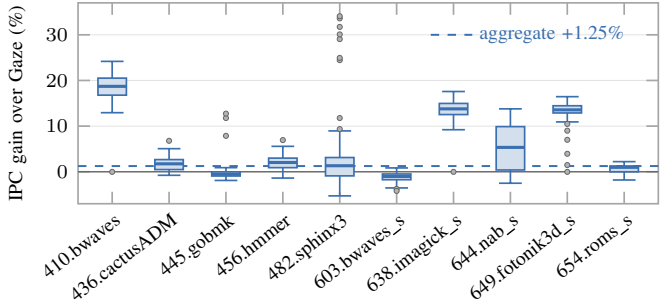
\begin{figure}[tb]
\centering
\resizebox{\columnwidth}{!}{
\begin{tikzpicture}
\begin{axis}[
  hpcafig,
  width=\columnwidth, height=4.1cm,
  ylabel={IPC gain over Gaze (\%)},
  ymin=-7, ymax=37,
  ytick={0,10,20,30},
  xmin=0.4, xmax=10.6,
  xtick={1,...,10},
  xticklabels={410.bwaves,436.cactusADM,445.gobmk,456.hmmer,482.sphinx3,
               603.bwaves\_s,638.imagick\_s,644.nab\_s,649.fotonik3d\_s,654.roms\_s},
  xticklabel style={font=\scriptsize, rotate=38, anchor=north east, yshift=1pt, xshift=2pt},
  boxplot/draw direction=y,
  boxplot/box extend=0.5,
  every boxplot/.style={
    draw=figblue, fill=figblue!22, line width=0.6pt, solid,
    mark=*, mark size=0.9pt,
    mark options={fill=gray!60, draw=figgray, line width=0.3pt, solid},
    /pgfplots/boxplot/every median/.style={figblue, line width=1.1pt, solid},
  },
  clip=false,
]
\addplot[figgray, line width=0.5pt, domain=0.4:10.6] {0};
\addplot[figblue, dashed, line width=0.7pt, domain=0.4:10.6] {1.25};
\draw[figblue, dashed, line width=0.7pt] (axis cs:6.6,30) -- (axis cs: 7.33,30);
\node[font=\scriptsize, text=figblue, anchor=west, inner sep=2pt] at (axis cs:7.35,30) {aggregate +1.25\%};
\addplot+[boxplot prepared={draw position=1, lower whisker=12.94, lower quartile=16.78, median=18.71, upper quartile=20.51, upper whisker=24.18}]
  coordinates {(1,-0.02)};
\addplot+[boxplot prepared={draw position=2, lower whisker=-0.77, lower quartile=0.48, median=1.74, upper quartile=2.67, upper whisker=5.06}]
  coordinates {(2,6.78)};
\addplot+[boxplot prepared={draw position=3, lower whisker=-1.90, lower quartile=-0.93, median=-0.55, upper quartile=-0.03, upper whisker=0.90}]
  coordinates {(3,7.87) (3,11.77) (3,12.73)};
\addplot+[boxplot prepared={draw position=4, lower whisker=-1.38, lower quartile=0.90, median=2.03, upper quartile=2.99, upper whisker=5.57}]
  coordinates {(4,6.94)};
\addplot+[boxplot prepared={draw position=5, lower whisker=-5.28, lower quartile=-0.90, median=1.32, upper quartile=3.12, upper whisker=8.95}]
  coordinates {(5,9.35) (5,11.77) (5,24.45) (5,24.94) (5,29.06) (5,30.09) (5,31.73) (5,33.60) (5,34.08)};
\addplot+[boxplot prepared={draw position=6, lower whisker=-3.54, lower quartile=-1.74, median=-1.03, upper quartile=-0.48, upper whisker=0.84}]
  coordinates {(6,-3.75) (6,-4.20)};
\addplot+[boxplot prepared={draw position=7, lower whisker=9.21, lower quartile=12.53, median=13.78, upper quartile=14.97, upper whisker=17.58}]
  coordinates {(7,-0.02)};
\addplot+[boxplot prepared={draw position=8, lower whisker=-2.51, lower quartile=0.39, median=5.34, upper quartile=9.89, upper whisker=13.78}]
  coordinates {};
\addplot+[boxplot prepared={draw position=9, lower whisker=10.92, lower quartile=12.88, median=13.56, upper quartile=14.46, upper whisker=16.45}]
  coordinates {(9,-0.02) (9,1.47) (9,3.94) (9,7.00) (9,9.00) (9,10.48)};
\addplot+[boxplot prepared={draw position=10, lower whisker=-1.80, lower quartile=0.00, median=0.97, upper quartile=1.26, upper whisker=2.22}]
  coordinates {};
\end{axis}
\end{tikzpicture}}
\caption{Per-trace IPC improvement of sliding-window UCB over always-Gaze under executed-performance feedback. UCB sees only the chosen policy's IPC; the dashed line marks its 1.25\% aggregate gain.}
\label{fig:two_policy}
\end{figure}

Figure~\ref{fig:two_policy} breaks the UCB result down by trace. The gain is concentrated in traces with persistent winner runs; we only include these traces in the figure. Traces with frequent Berti/Gaze ties or rapid alternation leave less structure for a history-only rule to exploit.

This result is encouraging but also exposes the tier's basic cost. UCB can learn that the inactive policy has become better only by giving that policy real control for a window. If the estimate is stale, waiting costs performance; if the PSB explores too aggressively, sampling costs performance. The winner stream is therefore predictable enough for adaptation, but executed-performance feedback pays for information by occasionally running the policy it suspects may be wrong.

\emph{Key takeaway: executed-performance feedback recovers 73.4\% of the pairwise gap with minimal interface changes, but must sometimes run the policy it aims to avoid.}


\subsection{Compact Confidence Signals}

The next question is whether the PSB can avoid that sampling cost by using information from the demand stream. The most compact version asks each prefetcher for a scalar confidence signal: a Berti delta-coverage score and a Gaze footprint-recurrence score. This is the obvious low-cost demand-stream selector, but it is also a lossy translation problem. The two scores are not expressed in a common unit; one summarizes repeated per-PC deltas, while the other summarizes recurring spatial footprints.

We test the direct comparison and several calibrations on the 20\% prefix data from the held-out benchmarks. Raw comparison selects the larger confidence value. Calibration fits models using the two scores together with their difference, ratio, product, and absolute difference. A second formulation predicts
\begin{equation}
\Delta IPC(w)=IPC_{\mathit{Berti}}(w)-IPC_{\mathit{Gaze}}(w)
\end{equation}
and selects from its sign.

\begin{table}[tb]
\centering
\caption{Confidence calibration on the 20\% prefix dataset.}
\label{tab:calibration}
\begin{tabular}{lcc}
\toprule
Method & Mean loss & Accuracy \\
\midrule
Always-Gaze & 1.237\% & 59.5\% \\
Raw confidence compare & 0.818\% & 50.3\% \\
Linear IPC-delta scaling & 0.780\% & 55.0\% \\
Best nonlinear model & 0.757\% & 50.3\% \\
Address-feature tree & 0.465\% & 72.6\% \\
\bottomrule
\end{tabular}
\end{table}

Table~\ref{tab:calibration} makes this a negative result. Raw confidence comparison reaches 0.818\% mean loss on held-out benchmarks. Linear rescaling improves that to 0.780\%, and the best nonlinear confidence model reaches 0.757\%, but all remain well behind the address-feature tree at 0.465\%. The IPC-delta regression has negative held-out $R^2$, meaning that on programs the model never saw during training, it predicts the size of the Berti/Gaze IPC difference worse than simply guessing the average. The confidence scores contain useful mechanism-local signals, but collapsing the demand stream to two scalars discards interleaving, density, recurrence, and volume context that the selector needs.

\emph{Key takeaway: mechanism-local confidence scores help, but two scalars discard too much demand-stream context to drive the selector.}


\subsection{Demand-Flow Tree}

The main practical PSB uses the richer demand-stream representation. During the sampling prefix, the passive monitor watches demand PCs and addresses and materializes six scalar feature values. These counters cover the regularity Berti exploits, the footprint recurrence Gaze exploits, and the mixing and concentration of the access stream across PCs and regions. They are computed before prefetch requests perturb the cache hierarchy, so the same feature vector is available regardless of which prefetcher is currently active.

The frozen tree gives this signal a small hardware interpretation. Offline training labels each training window with the better simulated policy and learns a depth-three tree. At runtime, the PSB traverses that tree: each internal node compares one of the six feature values with a stored threshold, takes the left or right child, and stops at a leaf that encodes Berti or Gaze. The unpruned tree has 15 entries: seven compare nodes and eight policy leaves. Inference therefore takes at most three threshold comparisons once per 200K-instruction window, not one prediction per memory reference.

With a 20\% prefix, the demand-flow tree reaches 0.465\% mean loss and 72.6\% accuracy on the 10 held-out benchmarks. Always-Gaze loses 1.237\% on the same held-out windows, so the tree recovers 62.4\% of the pairwise-oracle gap left by the static choice. Its aggregate IPC gain over always-Gaze on the held-out windows is 0.64\% (compared to UCB's aggregate gain of 0.58\% on the same held-out windows).

\emph{Key takeaway: a passive six-feature, depth-three tree recovers 62.4\% of the pairwise gap---higher accuracy than UCB at a slight loss cost---without running the inactive prefetcher.}


\subsection{Sampling-Prefix Sensitivity}

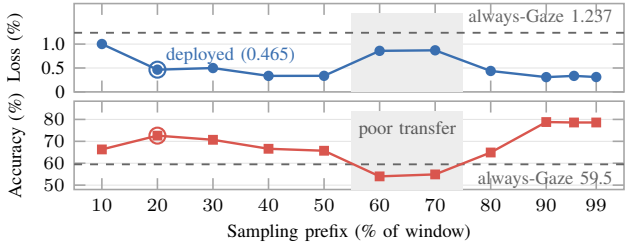
\begin{figure}[tb]
\centering
\begin{tikzpicture}
\begin{groupplot}[
  group style={group size=1 by 2, vertical sep=3.5pt, x descriptions at=edge bottom},
  hpcafig,
  width=\columnwidth, height=2.75cm,
  xmin=5, xmax=104,
  xtick={10,20,30,40,50,60,70,80,90,99},
  enlarge y limits={upper, value=0.12},
  clip=false,
]
\nextgroupplot[
  ylabel={Loss (\%)},
  ymin=0, ymax=1.65,
  ytick={0,0.5,1.0},
  yticklabels={0,0.5,1.0},
]
\path[fill=gray!14] (axis cs:55,0) rectangle (axis cs:75,1.65);
\addplot[figgray, dashed, line width=0.7pt, domain=5:104] {1.237};
\node[font=\scriptsize, text=figgray, anchor=south east, inner sep=2pt] at (axis cs:103,1.26) {always-Gaze 1.237};
\addplot[figblue, mark=*, mark size=1.5pt, mark options={fill=figblue}]
  coordinates {(10,1.002)(20,0.465)(30,0.499)(40,0.335)(50,0.336)(60,0.859)(70,0.870)(80,0.438)(90,0.309)(95,0.335)(99,0.311)};
\addplot[figblue, mark=o, mark size=3pt, only marks, line width=0.8pt] coordinates {(20,0.465)};
\node[font=\scriptsize, text=figblue, anchor=south west, inner sep=1pt] at (axis cs:21,0.52) {deployed (0.465)};
\nextgroupplot[
  ylabel={Accuracy (\%)},
  xlabel={Sampling prefix (\% of window)},
  xlabel style={yshift=2pt},
  ymin=48, ymax=84,
  ytick={50,60,70,80},
]
\path[fill=gray!14] (axis cs:55,48) rectangle (axis cs:75,84);
\node[font=\scriptsize, text=gray!130, anchor=north] at (axis cs:65,83.5) {poor transfer};
\addplot[figgray, dashed, line width=0.7pt, domain=5:104] {59.5};
\node[font=\scriptsize, text=figgray, anchor=north east, inner sep=2pt] at (axis cs:103,58.8) {always-Gaze 59.5};
\addplot[figred, mark=square*, mark size=1.4pt, mark options={fill=figred}]
  coordinates {(10,66.3)(20,72.6)(30,70.7)(40,66.6)(50,65.7)(60,54.0)(70,54.9)(80,64.9)(90,78.8)(95,78.6)(99,78.6)};
\addplot[figred, mark=o, mark size=3pt, only marks, line width=0.8pt] coordinates {(20,72.6)};
\end{groupplot}
\end{tikzpicture}
\caption{Held-out prefix sweep. The circled 20\% point denotes the selected practical design. The shaded 60--70\% region reflects cross-benchmark transfer degradation despite the larger observation set. High-prefix points are diagnostic because little of the window remains after prediction.}
\label{fig:sampling_tradeoff}
\end{figure}

A demand-stream PSB cannot choose before it has observed any demand references. The prefix is the fraction of the current 200K-instruction window used for observation before the PSB selects a policy for the rest of that same window. A longer prefix gives the monitor more stream context, but leaves less of the window to benefit from the decision. Figure~\ref{fig:sampling_tradeoff} sweeps that trade-off on the held-out benchmarks.

The sweep is not monotonic: more observation does not reliably help. A 10\% prefix provides too little stream context. Prefixes from 20--50\% retain most of the opportunity, while 60--70\% transfer poorly to the held-out benchmarks. The 90--99\% rows show that the features can recover a strong signal when nearly the whole window is visible, but those points are diagnostic rather than attractive same-window controllers because little execution remains after the decision. We select 20\% because it is the best short prefix in this split and preserves a substantial post-decision interval.

Two effects can explain the 60--70\% degradation. A program can change behavior around the sampling boundary, so the prefix and the remainder of the window favor different policies. Alternatively, a longer prefix can move a feature value across a threshold learned from other benchmarks, improving the description of the current interval while hurting transfer through that stored threshold. These traces do not distinguish the causes; multiple benchmark splits and explicit feature-drift measurements would be needed to separate phase placement from threshold sensitivity.

\emph{Key takeaway: more observation is not automatically better: the 20\% prefix gives enough context while leaving most of the window for the chosen policy.}


\subsection{Feature, Depth, and Residence Sensitivity}

Table~\ref{tab:tree_sensitivity} asks whether the demand-flow tree needs all of its inputs and whether a larger tree helps. ``Mechanism'' keeps the Berti- and Gaze-oriented summaries. ``+ interleaving'' adds PC/region mixing. ``All'' uses the complete six-feature vector.

\begin{table}[tb]
\centering
\caption{Feature and depth sensitivity for the 20\% demand selector.}
\label{tab:tree_sensitivity}
\begin{tabular}{lccc}
\toprule
Feature set & Depth & Mean loss & Accuracy \\
\midrule
Mechanism & 3 & 0.777\% & 52.6\% \\
+ interleaving & 3 & 0.821\% & 65.2\% \\
All & 1 & 1.288\% & 55.9\% \\
All & 3 & 0.465\% & 72.6\% \\
All & 5 & 0.642\% & 61.5\% \\
\bottomrule
\end{tabular}
\end{table}

Mechanism-only features reach 0.777\% loss and 52.6\% accuracy. Adding interleaving raises accuracy to 65.2\% but does not reduce loss, which means the extra correct decisions are concentrated in windows where Berti and Gaze were close, while some higher-margin mistakes remain. The complete depth-three tree gives the best joint result. Depth one is too small to separate the relevant stream cases and loses 1.288\%; depth five has more capacity but raises held-out loss to 0.642\%, consistent with memorizing training-program details that do not transfer. The hardware-friendly point is also the best evaluated point: the complete six-feature vector and at most three comparisons.

Switching frequency determines how often the state-management question in Section~\ref{sec:selector_design} matters. The unconstrained tree requests 13.9 switches per 100-window test trace. A minimum residence time adds hysteresis without retraining the classifier: once the PSB chooses a policy, later predictions are ignored until the hold interval expires. A two-window hold reduces switching by 36\%, to 8.9 per trace, while loss rises from 0.465\% to 0.570\%. Five- and ten-window holds reduce activity further but increase loss above 0.7\%. A 20-window hold reduces switching by 88\% and raises loss to 0.850\%. Thus, modest hysteresis can reduce state-management activity, while long holds discard much of the selector's benefit.

\begin{table}[tb]
\centering
\caption{Minimum-residence sensitivity for the 20\% demand tree.}
\label{tab:hold_sensitivity}
\begin{tabular}{rccc}
\toprule
Hold & Mean loss & Accuracy & Sw./trace \\
\midrule
1 & 0.465\% & 72.6\% & 13.9 \\
2 & 0.570\% & 69.7\% & 8.9 \\
5 & 0.721\% & 68.4\% & 5.4 \\
10 & 0.758\% & 68.2\% & 3.3 \\
20 & 0.850\% & 65.9\% & 1.6 \\
\bottomrule
\end{tabular}
\end{table}

These sensitivity studies still do not synthesize monitor area, monitor energy, finite-table aliasing, or cycle-level switch latency. Because switches reset prefetcher metadata rather than transferring state between Berti and Gaze, the residence sweep bounds how often reset and relearning costs can occur. It does not convert those events into cycles, joules, or storage overhead, so choosing a final hold interval requires an implementation study.

\emph{Key takeaway: the selector is small but not arbitrary---it needs all six features at depth three, and modest residence hysteresis trims switching at a controlled loss cost.}


\subsection{Policy-Pair Diversity}

The previous results use the selected Berti/Gaze pair from Section~\ref{sec:compression}. We also apply the same 20\% feature construction and depth-three tree to every pair among the eight configurations to check whether selector value comes from the PSB alone or from choosing a pair with real headroom. Table~\ref{tab:pairs} reports representative same-family (i.e. both configs include the same L1D prefetcher) and Berti/Gaze cases.

\begin{table}[tb]
\centering
\caption{Representative pairwise diagnostics for the 20\% address tree. Ent.\ denotes Entangling and Mock.\ denotes Mockingjay.}
\label{tab:pairs}
\begin{tabular}{lcc}
\toprule
Pair & Loss & Mean gap \\
\midrule
Berti/BARCA/PACIPV vs.\ Berti/Ent./PACIPV & 0.029\% & 0.080\% \\
Berti/Ent./PACIPV vs.\ Berti/BARCA/Mock. & 0.033\% & 0.104\% \\
Berti/Ent./PACIPV vs.\ Berti/Ent./Mock. & 0.013\% & 0.027\% \\
Berti/BARCA/Mock. vs.\ Berti/Ent./Mock. & 0.025\% & 0.080\% \\
Best Berti/Gaze & 0.422\% & 2.007\% \\
\bottomrule
\end{tabular}
\end{table}

Same-family pairs have mean gaps near or below 0.1\% and selector loss below 0.04\%, leaving little pairwise opportunity. Berti/Gaze pairs present larger gaps near 2\%, so the selector has a meaningful decision to make. The best-performing diagnostic Berti/Gaze pair has 2.007\% mean gap and 0.422\% selector loss. The headline result remains 0.465\% because the primary pair and held-out protocol were chosen in advance by the Section~\ref{sec:compression} criteria, rather than by selecting the best pair after this sweep.

\begin{figure}[t]
\centering
\begin{tikzpicture}
\begin{axis}[
  name=gapL,
  hpcafig,
  width=0.46\columnwidth, height=3.25cm,
  scale only axis=false,
  xmin=0, xmax=0.12,
  ymin=-0.25, ymax=1.25,
  xtick={0,0.05,0.10},
  xticklabels={0,0.05,0.10},
  ytick={0,1},
  yticklabels={{Berti/Gaze},{same L1D}},
  yticklabel style={font=\scriptsize, align=right},
  axis x line*=bottom, axis y line*=left,
]
\addplot[only marks, mark=*, mark size=2.0pt,
  mark options={fill=gray!55, draw=figgray, line width=0.4pt}]
coordinates {
  (0.0153,1.04) (0.0156,0.95) (0.0244,1.08) (0.0266,0.91)
  (0.0660,1.02) (0.0674,0.93) (0.0788,1.09) (0.0818,0.96)
  (0.0836,1.03) (0.0842,0.89) (0.1039,1.07) (0.1050,0.95)
};
\end{axis}

\begin{axis}[
  name=gapR,
  hpcafig,
  at={($(gapL.east)+(5.5mm,0)$)}, anchor=west,
  width=0.52\columnwidth, height=3.25cm,
  scale only axis=false,
  xmin=1.968, xmax=2.012,
  ymin=-0.25, ymax=1.25,
  xtick={1.97,1.98,1.99,2.00,2.01},
  xticklabels={1.97,1.98,1.99,2.00,2.01},
  ytick={0,1},
  yticklabels={},
  axis x line*=bottom, axis y line*=left,
]
\addplot[only marks, mark=triangle*, mark size=2.3pt,
  mark options={fill=figblue!85, draw=figblue, line width=0.4pt}]
coordinates {
  (1.9735,0.06) (1.9738,-0.04) (1.9760,0.10) (1.9768,-0.09)
  (1.9804,0.05) (1.9806,-0.03) (1.9814,0.11) (1.9820,-0.10)
  (1.9834,0.02) (1.9841,-0.06) (1.9892,0.08) (1.9897,-0.01)
  (2.0011,0.07) (2.0014,-0.08) (2.0068,0.04) (2.0070,-0.04)
};
\end{axis}

\path (gapL.south east) -- (gapR.south west) coordinate[midway] (brk);
\draw[gray!70, semithick]
  ($(brk)+(-1.6mm,-1.1mm)$) -- ($(brk)+(-0.4mm,1.1mm)$);
\draw[gray!70, semithick]
  ($(brk)+(0.4mm,-1.1mm)$) -- ($(brk)+(1.6mm,1.1mm)$);

\path (gapL.south) -- (gapR.south) coordinate[midway] (xlab);
\node[font=\scriptsize, anchor=north, yshift=-4.5mm] at (xlab)
  {Mean pairwise IPC gap (\%)};
\end{tikzpicture}
\caption{Mean IPC-gap distribution for all 28 policy pairs. The axis break separates 12 same-L1D pairs from 16 Berti/Gaze pairs; vertical jitter only separates overlapping points.}
\label{fig:pairwise_gap}
\end{figure}
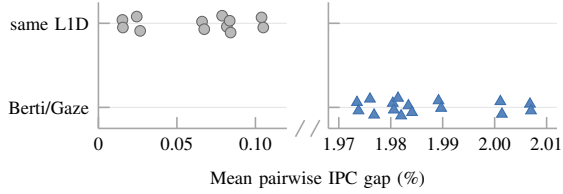

The pairwise check reinforces the two-stage design method. First choose a candidate subset using oracle coverage, complementarity, and implementation scope; then evaluate how much of that subset's gap an implementable PSB can recover. A selector can look accurate on a pair with many ties, and it can look low-loss on a pair with no useful headroom. Pair selection and selector evaluation have to be interpreted together.

\emph{Key takeaway: selector quality matters only when the pair exposes real complementary headroom, which is why Berti/Gaze beats same-family pairs.}


\subsection{Information Boundary and Counterfactual Upper Bound}
\label{sec:ipu}

The practical tiers have different failure modes, but they share the same blind spot: neither directly observes the inactive prefetcher's consequences. Executed-performance feedback observes Berti only after running Berti and Gaze only after running Gaze. Demand-stream monitoring observes the program-side access pattern without running the inactive engine, but it still cannot see whether that engine's requests would have arrived in time, displaced useful lines, consumed bandwidth, or occupied MSHRs. Two windows can therefore have similar demand-flow counters and different correct policies because the hidden prefetcher consequences differ.

We bound the value of that missing signal with an ideal counterfactual observer. The observer reports which prefetcher won in a completed window without allowing the inactive prefetcher to issue requests or perturb cache state. Even this ideal signal cannot predict the future: a winner observed for window $w$ can first affect the decision for window $w+1$. The best simple sequence rule is therefore repeat-last, which uses the most recent observed winner as the next decision.

\begin{table}[tb]
\centering
\caption{Idealized winner-history prediction. The demand-only result provides a reference without counterfactual information.}
\label{tab:ipu_results}
\footnotesize
\begin{tabular}{@{}lcc@{}}
\toprule
Evaluation & Mean loss & Accuracy \\
\midrule
Demand, fixed split & 0.465\% & 72.60\% \\
\midrule
Winner, fixed split & 0.320\% & 74.50\% \\
Winner, leave-one-out & 0.256\% & 74.35\% \\
Winner, 100-split mean & 0.263\% & 73.68\% \\
\bottomrule
\end{tabular}
\end{table}

Table~\ref{tab:ipu_results} shows the upper bound. On the same fixed 39/10 split as the demand tree, exact winner history reduces mean loss from 0.465\% to 0.320\%, a 0.145-point improvement. Leave-one-benchmark-out and 100 random 39/10 splits give similar losses, indicating that the result comes from adjacent-window winner persistence rather than a lucky split. The improvement is real, but it is much smaller than the 0.772-point gain already obtained by passive demand monitoring over always-Gaze on the held-out set.

The bound is also fragile. A second window of observation delay raises loss to 0.464\%, effectively matching the 0.465\% demand-tree result; 10\% winner-observation error raises loss to 0.458\%, again removing most of the advantage. Five-percent error retains a measurable benefit at 0.387\%. Thus counterfactual support is attractive only if a practical observer can deliver a near-exact winner signal within one window, or if suitable introspection hardware already exists for another purpose. For the evaluated workloads, passive demand-stream monitoring remains the preferred standalone design point.

\emph{Key takeaway: ideal counterfactual winner information gives a real but small gain that vanishes under modest delay or error, leaving passive demand monitoring as the preferred PSB.}


\section{Related Work}
\label{sec:related}

\paragraph{Learned policies and bounded selection.}
Learned prefetchers establish that address streams contain predictable structure. Hashemi et al.\ formulate prefetching as sequence prediction~\cite{hashemi2018learning}; Voyager separates page and offset behavior~\cite{shi2021hierarchical}; and DART distills an attention model into hardware-oriented tables~\cite{zhang2024attention}. Other learned microarchitectural mechanisms become practical by restricting their state or action space, including perceptron branch prediction~\cite{jimenez2001perceptron}, learned memory scheduling~\cite{ipek2008memory}, learned cache replacement~\cite{jain2016hawkeye,shi2019glider}, and Pythia's structured reinforcement-learning formulation~\cite{bera2021pythia}. We use learning for a narrower task: Berti and Gaze retain their expert-designed request generators, while a once-per-window selector produces one policy bit off the per-access critical path.

This separation changes both the learning target and the implementation burden. A learned prefetcher must decide which address to request, often at access granularity, and its mistakes directly consume cache and memory resources. Our model never generates an address. It chooses between two independently validated mechanisms, executes only the selected engine, and leaves the request path unchanged. Offline paired runs provide labels during model construction, but the deployed tree receives neither paired IPC nor inactive-policy events. The result is closer to mechanism selection than policy synthesis: model capacity is deliberately limited, and the principal question is whether passive demand behavior contains enough information to choose the existing mechanism.

Our candidates span distinct parts of the memory hierarchy. Berti~\cite{navarro2022berti} and Gaze~\cite{chen2025gaze} are mechanically different L1D prefetchers; Entangling~\cite{ros2021cost} and BARCA~\cite{gratz2020barca} target instruction delivery; and Mockingjay~\cite{shah2022effective} and PACIPV~\cite{mostofi2025light} improve replacement. Rather than assume a candidate pair, we first evaluate all eight L1D$\times$L1I$\times$L2 compositions and then reduce the action space according to oracle coverage, complementarity, and switching scope.

\paragraph{Adaptive hardware and prefetcher selection.}
Processors commonly choose among bounded alternatives. DIP and DRRIP use set dueling~\cite{qureshi2007adaptive,jaleel2010high}; tournament branch predictors select a component~\cite{mcfarling1993combining,mittal2019survey}; utility-based partitioning estimates cache benefit with auxiliary tags~\cite{qureshi2006utility}; and Micro-Armed Bandit applies UCB-family rules to microarchitectural actions~\cite{gerogiannis2023bandit}. These mechanisms can observe an eventual outcome or confine sampling to a limited structure. Selecting a complete prefetcher is harder: the inactive engine would have changed cache contents, bandwidth, MSHR occupancy, and future request timing, so its outcome is not locally observable.

Alcorta et al.\ study lightweight ML-based prefetcher selection and the effects of decision granularity~\cite{alcorta2023lightweight,alcorta2024characterizing}. Feedback-directed control adjusts aggressiveness~\cite{srinath2007feedback}, coordinated control manages interference among active prefetchers~\cite{ebrahimi2009coordinated}, and Sandbox Prefetching, Best-Offset, and IPCP evaluate or route among integrated candidates~\cite{pugsley2014sandbox,michaud2016best,pakalapati2020bouquet}. Our selector instead chooses one unmodified L1D engine while the other remains inactive. Its demand monitor is read-only and action independent; consequently, it avoids perturbing the hierarchy but must infer the inactive engine's utility.

The distinction is especially important for feedback. Set dueling assigns a small number of cache sets to competing policies, and a tournament predictor observes a correctness outcome for the component that supplied a prediction. An inactive prefetcher has no analogous local outcome: suppressing its requests also removes their timeliness, pollution, and contention effects. Our UCB replay makes this limitation explicit by updating only the chosen arm, while the demand selector avoids exploration by observing an action-independent stream. Evaluating both mechanisms on the same policy pair separates gains due to temporal persistence from gains due to a richer program-side signal.

\paragraph{Phase prediction and counterfactual observation.}
SimPoint and working-set signatures show that programs revisit distinguishable phases~\cite{sherwood2002simpoint,dhodapkar2002workingset}, and later work uses learned models to forecast behavior or tune hardware~\cite{lozano2022learning,ortega2020intelligent,wadle2025sahm}. Our design follows this coarse-grained principle: offline replay supplies paired IPC labels, whereas runtime hardware retains PC/region summaries and a small decision tree. The prefix sweep further shows that longer observation can cross a phase boundary and need not improve transfer.

Phase detection and policy selection are nevertheless different problems. A stable signature can identify that execution has revisited a region without revealing which of two prefetchers will convert that behavior into timely, nonpolluting requests. Conversely, a selector can tolerate imperfect phase boundaries when errors occur in low-gap windows. We therefore evaluate both decision accuracy and IPC loss, freeze the learned boundary before benchmark-disjoint testing, and retain chronological order for replay. This protocol tests transfer of a compact decision rule rather than memorization of trace identifiers.

Finally, the IPU offers a programmable substrate for introspection and in-field A/B evaluation~\cite{mcdougall2025ipu}. We use an idealized observer to bound the value of such counterfactual information, not to claim an implementation. Its delay and error checks quantify the signal quality required to outperform passive observation; realizing that signal within area, energy, verification, and security constraints remains future work.

\section{Conclusion}
This paper argues that runtime adaptation should start by asking which policies
are worth selecting, rather than assuming every available mechanism belongs in
the action space. In our memory-system study, Gaze/Entangling/Mockingjay is the
BGSP by phase-level oracle-win frequency, yet remains 1.33\% from the
eight-configuration oracle on average and exposes a concentrated phase tail. That
opportunity compresses to a Berti/Gaze L1D-prefetcher choice within 0.039\%
aggregate IPC of the full oracle. A passive PSB built from either
executed-performance feedback or passive demand monitoring captures much of that
opportunity, recovering 62.4\% to 73.4\% of the pairwise oracle gap without
executing or emulating the inactive prefetcher. The counterfactual study bounds
the rest: exact winner history reduces the remaining loss, but one window of
delay or 10\% observation error returns performance to the demand-only level.
Passive executed-performance feedback or demand monitoring are therefore the right design points for this
policy set, with counterfactual support compelling only if a near-exact,
one-window-latency observer can be built or amortized by other hardware. More
broadly, this framing applies wherever the design choice is a policy rather than a
structural size, since choosing among predictors, prefetchers, replacement rules,
or scheduling heuristics is more plausible than dynamically turning a 4-wide core
into an 8-wide core. The next step is to extend this binary selector to small
multi-policy action spaces under the same discipline: compress candidates by
oracle coverage and complementarity, expose only an interface the machine can
afford, and judge future counterfactual monitors by whether their latency and
error actually beat passive demand observation.

\clearpage
\bibliographystyle{IEEEtranS}
\bibliography{refs}

\end{document}